\documentclass[onecolumn,aps,prd,showpacs,superscriptaddress,nofootinbib,amsmath,amssymb,floats,floatfix,showkeys,notitlepage,longbibliography, 10pt]{revtex4-1}

\usepackage[english]{babel}
\usepackage{comment}
\usepackage{xcolor}
\usepackage{graphicx}
\usepackage{subfigure}
\usepackage{palatino}
\usepackage[commandnameprefix=always]{changes}
\usepackage{hyperref}
\hypersetup{colorlinks=true,linkcolor=blue,urlcolor=blue,citecolor=blue}
\usepackage[toc,page]{appendix}
\usepackage[normalem]{ulem}
\usepackage{orcidlink}
\usepackage{lipsum}
\usepackage{graphicx}
\usepackage{subfigure}
\usepackage{palatino}
\usepackage{sans}
\usepackage{adjustbox}
\usepackage{latexsym}
\usepackage{amsmath}
\usepackage{amssymb}
\usepackage{amsfonts}
\usepackage{dcolumn}
\usepackage{bm}
\usepackage{tikz}
\usepackage{bigints}
\usepackage{array,tabularx,multirow,booktabs}
\usepackage[tracking=true]{microtype}
\SetTracking{}{500}
\SetTracking{encoding={*}, shape=sc}{40}
\UseRawInputEncoding 
\allowdisplaybreaks
\usepackage{adjustbox}
\usepackage{latexsym}
\usepackage{amsmath}
\usepackage{amssymb}
\usepackage{amsfonts}
\usepackage{dcolumn}
\usepackage{bm}
\usepackage{tikz}
\usepackage{bigints}
\usepackage{array,tabularx,multirow,booktabs}
\usepackage[tracking=true]{microtype}
\usepackage{color}
\UseRawInputEncoding 
\allowdisplaybreaks

\begin{document} 

\title{Barotropic Equation of State and Nonlinear Electrodynamics in Dynamical Black Hole Spacetimes}

\author{Vitalii Vertogradov}
\email{vdvertogradov@gmail.com}
\affiliation{Physics department, Herzen state Pedagogical University of Russia,
48 Moika Emb., Saint Petersburg 191186, Russia} 
\affiliation{Center for Theoretical Physics, Khazar University, 41 Mehseti Street, Baku, AZ-1096, Azerbaijan.}
\affiliation{SPB branch of SAO RAS, 65 Pulkovskoe Rd, Saint Petersburg
196140, Russia}

\author{Dmitriy Kudryavtcev }
\email{kudryavtsiev33@gmail.com}
\affiliation{Department of High Energy and Elementary Particles Physics, Saint Petersburg State University, University Emb. 7/9, Saint Petersburg, 199034, Russia}

\begin{abstract}
Models of black holes that differ from idealized vacuum and electrovacuum solutions of Einstein's equations often contain parameters whose physical interpretation is unclear. However, to propose a black hole model for experimental verification, we must clearly understand which parameters describe the black hole and what physical constraints can be imposed on each parameter. When considering dynamical black holes with a barotropic equation of state $P = \alpha \rho$, an additional integration constant arises, the nature of which remains unclear except for a few specific values of the equation-of-state coefficient $\alpha$. Nevertheless, when examining solutions such as Husain's or Kiselev's, we can observe remarkable effects related to black hole evaporation, which are associated with the violation of null energy conditions. 

However, the main issue lies in the fact that while violations of energy conditions may occur in nature, the ambiguity in the values of parameters describing black holes makes it difficult to determine whether such violations result from natural physical processes or are purely mathematical artifacts that should be excluded from consideration. Moreover, energy conditions play a crucial role in the evolution of a black hole's shadow — a characteristic that can potentially be observed experimentally. 

In this article, we elucidate the parameter arising in Husain's and Kiselev's solutions using nonlinear electrodynamics. We demonstrate that for physically relevant equations of state, the additional parameter represents a combination of electric and magnetic charges.

\end{abstract}

\date{\today}



\maketitle
\section{Introduction}

Following the detection of black hole shadows by the Event Horizon Telescope collaboration in the centers of galaxies M87~\cite{bib:ehtm1} and the Milky Way~\cite{bib:ehtc3}, black holes have become one of the central objects of modern theoretical physics. In this regard, it is crucial to find solutions to Einstein's equations that describe realistic astrophysical black holes. Among the most well-known solutions in general relativity—such as the Schwarzschild, Reissner-Nordström, Kerr, and Kerr-Newman metrics—are idealizations that are useful primarily for qualitative investigations of black hole properties.

Beyond these solutions, there exist numerous other solutions to Einstein's equations describing black holes~\cite{bib:ali2025cqg, bib:dym, bib:bardeen, bib:rincon1, bib:rincon2, bib:hay, bib:daniil2025barionic, bib:vertogradov2025podu, bib:ali2025new, bib:husain1996exact, bib:kiselev, bib:tur1, bib:tur2, bib:tur3, bib:misyura2024podu, bib:bh1, bib:bh2, bib:vertogradov2024grg, bib:baptista2024}. These solutions differ from the classical Schwarzschild, Kerr, Reissner-Nordström, and Kerr-Newman metrics and, as such, are not vacuum or electro-vacuum solutions of general relativity. Consequently, these solutions typically contain two integration constants: one associated with the black hole's mass and another whose physical nature often remains unknown.

In addition to shadow observations, black holes can also be detected in the optical range under specific conditions. For instance, if a black hole moves slowly through a cloud of gas, such as hydrogen gas, the accretion processes onto the black hole can be studied and observed~\cite{bib:beskin}. However, these processes must exhibit distinct features compared to similar phenomena involving neutron stars. This highlights the need to investigate black hole solutions in general relativity that go beyond vacuum and electro-vacuum solutions. As noted earlier, most alternative black hole models include an additional parameter alongside the mass, whose nature must be determined to impose constraints on its value.

Furthermore, when a black hole accretes matter, it can no longer be described by static or stationary metrics, as its mass continuously increases during the accretion process. Consequently, the spacetime around the black hole must be dynamic. One of the simplest models of dynamic black holes is the Vaidya black hole~\cite{bib:vaidya}, or radiating Schwarzschild black hole. However, like the Schwarzschild metric, it is an idealization. A more comprehensive approach involves considering field equations with a non-zero Type-I matter field on the right-hand side of Einstein's equations. This leads to the generalized spacetime metric~\cite{bib:vunk}:
\begin{equation}
ds^2 = -\left(1 - \frac{2M(v,r)}{r}\right)dv^2 + 2dvdr + r^2d\Omega^2,
\end{equation}
where $M(v,r)$ is the mass function.

One of the simplest equations of state is the barotropic equation of state:
\begin{equation}
P = \alpha \rho.
\end{equation}
In this case, solving Einstein's equations yields the Husain solution~\cite{bib:husain1996exact}, for which the mass function takes the form:
\begin{equation}
M(v,r) = M_0(v) + D(v)r^{1-2\alpha},
\end{equation}
where $M_0(v)$ represents the black hole's mass, while the physical nature of the function $D(v)$ remains unclear.

An important remark should be made regarding the parameter $\alpha$: in fact, due to the anisotropic nature of pressure in this model, the interpretation of $\alpha$ as a parameter of the barotropic equation of state is not entirely accurate. To derive a barotropic equation of state, it is common to consider the averaged pressure $\bar{P}$, defined as:
\begin{equation}
    \bar{P} = \frac{1}{3} \left( P_r + 2P_t \right),
\end{equation}
where $P_r$ and $P_t$ are the radial and tangential components of pressure, respectively. The barotropic equation of state then takes the form:
\begin{equation}
    \bar{P} = \omega \rho.
\end{equation}
It is straightforward to show that $\alpha$ and $\omega$ are related by:
\begin{equation}
    \alpha = \frac{1}{2} \left( 3\omega + 1 \right).
\end{equation}
Physically, the barotropic equation of state with $\omega \in [0, 1]$ corresponds to values of $\alpha \in \left[\frac{1}{2}, 2\right]$. In light of this, we will use the parameter $\alpha$ for computational simplicity, but restrict our analysis to the physically meaningful case of $\alpha > \frac{1}{2}$.

Understanding the nature of $D(v)$ is essential for several reasons:

- Clarifying the physical origin of $D(v)$ allows us to impose constraints on this function and assess whether the Husain solution can describe real black holes.
    
- The Husain solution plays a key role in studying gravitational collapse processes leading to regular black holes~\cite{bib:vertogradov2025podu, bib:ali2025new}.
    
- When analyzing the dynamical properties of the shadow of a Husain black hole, cases arise where null energy conditions are violated, resulting in a reduction of the shadow size during accretion. Understanding the physical nature of $D(v)$ will help determine whether such behavior is physically plausible or merely a mathematical artifact caused by the violation of energy conditions~\cite{bib:vertogradov2024horizon}.

In this paper, we attempt to elucidate the physical nature of $D(v)$ using nonlinear electrodynamics. We demonstrate that for certain equations of state where the parameter $\alpha \in \left(\frac{1}{2}, 1\right]$, the function $D(v)$ may represent a combination of electric and magnetic charges. Nonlinear electrodynamics~\cite{bib:nonlinear} gained widespread application in the context of black holes following the work~\cite{bib:garcia}, which demonstrated that the parameter $g$, preventing the formation of a singularity in the Bardeen metric, originates from nonlinear electrodynamics. Subsequent studies revealed that a regular center can form if and only if electric fields are absent, while the black hole is described by a magnetic monopole charge~\cite{bib:bronnikov}. When considering the gravitational collapse of a Husain black hole without assuming phase transitions in the matter, the inevitable outcome is the formation of a singularity. In light of this, in the present work, we examine not only the case of a magnetic field but also electric and electromagnetic fields.

The paper is organized as follows: Section 2 provides a detailed description of the Husain solution and its key properties. Section 3 derives the Einstein equations coupled with nonlinear electrodynamics. Section 4 investigates the physical nature of $D(v)$. Finally, Section 5 presents the conclusions.

Throughout the paper, we use geometrized units where $c = 8\pi G = 1$ and adopt the signature $(-+++)$.
\section{Husain Solution}

The most general form of a spherically symmetric dynamical spacetime can be expressed as:
\begin{equation}
ds^2 = -A^2(v,r)f(v,r)dv^2 + 2\varepsilon A(v,r)dvdr + r^2d\Omega^2,
\end{equation}
where $d\Omega^2 = d\theta^2 + \sin^2\theta d\varphi^2$ is the metric on the unit two-sphere, and $\varepsilon = \pm 1$ depends on the process under consideration. In what follows, we restrict our analysis to the accretion case, where $\varepsilon = +1$. The function $A(v,r) > 0$, and the mass function $M(v,r)$ is related to $f(v,r)$ via:
\begin{equation}
f(v,r) = 1 - \frac{2M(v,r)}{r}.
\end{equation}
We further consider a simplified model where $A(v,r) \equiv 1$. Consequently, the spacetime metric takes the form:
\begin{equation} \label{eq:metric}
ds^2 = -\left(1 - \frac{2M(v,r)}{r}\right)dv^2 + 2dvdr + r^2d\Omega^2.
\end{equation}
The Einstein field equations for this metric yield:
\begin{eqnarray} \label{eq:einstein}
\mu(v,r) &=& \frac{2\dot{M}(v)}{r^2}, \nonumber \\
\rho(v,r) &=& \frac{2M'(v,r)}{r^2}, \nonumber \\
P(v,r) &=& -\frac{M''(v,r)}{r},
\end{eqnarray}
where $\mu$ represents the radiation energy density, while $\rho$ and $P$ denote the energy density and pressure of matter, respectively. To solve this system of differential equations, an additional equation of state $P = P(\rho)$ must be introduced to characterize the properties of the matter.

Considering a barotropic equation of state:
\begin{equation}
P = \alpha \rho, \quad \alpha \in [-1,1], \quad \alpha \neq \frac{1}{2},
\end{equation}
leads to the solution of the Einstein equations known as the Husain solution~\cite{bib:husain1996exact}:
\begin{equation} \label{eq:husain}
M(v,r) = M_0(v) + D(v)r^{1-2\alpha}.
\end{equation}
Here, $M_0(v)$ can be associated with the black hole mass, while the physical interpretation of the parameter $D(v)$ remains unclear and will be discussed below.

The Husain solution is widely employed in the literature to describe gravitational collapse~\cite{bib:maombi, bib:vertogradov2016gc, bib:vertogradov2022ijmpa}, serving as the external solution for collapsing matter that leads to the formation of a regular black hole~\cite{bib:vertogradov2025podu, bib:ali2025new, bib:daniil2025barionic}. If the metric~\eqref{eq:metric} admits a homothetic Killing vector, it reduces to the Husain solution~\eqref{eq:husain}~\cite{bib:maharaj, bib:vertogradov2023mpla}.

The Husain solution must satisfy the weak energy conditions, which require that the energy density measured by any timelike observer remains non-negative. This leads to the following two cases:
\begin{enumerate}
\item If $\alpha < \frac{1}{2}$, the weak energy conditions require $D(v) > 0$.
\item If $\alpha > \frac{1}{2}$, the weak energy conditions require $D(v) < 0$.
\end{enumerate}

In the dynamical case, the event horizon cannot be defined. Instead, the boundary of the black hole is determined by the marginal trapping hypersurface, which serves as the apparent horizon. Initially, it was assumed that the apparent horizon is a spacelike hypersurface; however, the application of Hawking radiation suggests that it may also be timelike. In~\cite{bib:vertogradov2024horizon}, the relationship between the behavior of inner and outer apparent horizons is analyzed based on null energy conditions.

\begin{figure}[htbp] 
    \centering 
    \includegraphics[width=0.5\textwidth]{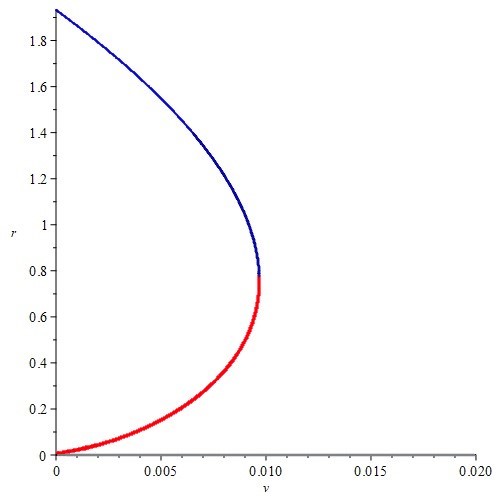} 
    \caption{\textit{The plot illustrates the apparent horizons for the Husain black hole with parameters } $\lambda = 1$, $\alpha = 4/5$, $M_0 = 1$, $D_0 = 0.1$. \textit{This case corresponds to the choice of functions given by Eq.}~\eqref{eq:vibor}, \textit{which violate the null energy conditions. As a result, the outer apparent horizon, despite the increasing mass, turns out to be a timelike hypersurface that shrinks over time (blue curve), while the inner horizon is also timelike but expands (red curve). At a certain moment, the two horizons merge and subsequently disappear. This peculiar dynamical behavior calls for a deeper understanding of the nature of the function } $D(v)$, \textit{in order to identify the physical processes that could give rise to such evolution of the horizons.}} 
    \label{fig:1} 
\end{figure}

The apparent horizon for the Husain solution is given by:
\begin{equation}
1 - \frac{2M_0(v)}{r_{ah}} - \frac{2D(v)}{r_{ah}^{2\alpha}} = 0,
\end{equation}
where $r = r_{ah}$ denotes the location of the apparent horizons. Notably, when $\alpha < 0$, the Husain black hole has a single apparent horizon. However, due to the loss of asymptotic flatness in this case, an outer horizon emerges, acting as a cosmological horizon. For $\alpha \in [0, \frac{1}{2}]$, the Husain black hole possesses one apparent horizon, while for $\alpha \in (\frac{1}{2}, 1]$, it exhibits two apparent horizons, whose dynamics depend on the energy conditions. If the null energy conditions hold throughout the spacetime, the outer apparent horizon is spacelike and expands, whereas the inner horizon is timelike and contracts, eventually leading to the Vaidya metric in the asymptotic limit. As an example of such behavior, Figure~\ref{fig:1} shows the evolution of apparent horizons when the isotropic energy conditions are violated. The red curve illustrates the behavior of the inner horizon, which is a spacelike hypersurface, while the blue curve represents the evolution of the outer timelike apparent horizon, which contracts over time. At a specific moment, these horizons merge and subsequently disappear. This behavior corresponds to the following choice of functions:
\begin{eqnarray} \label{eq:vibor}
M(v) &=& M_0 + \lambda v, \quad \lambda > 0, \nonumber \\
D(v) &=& -D_0 - 100\lambda v, \nonumber \\
\alpha &=& \frac{4}{5}.
\end{eqnarray}

The presence of two apparent horizons, analogous to the Reissner-Nordström solution, suggests that the nature of $D(v)$ might be linked to electric or magnetic charges arising from nonlinear electrodynamics. Motivated by this, we proceed under the assumptions $\alpha > \frac{1}{2}$ and $D(v) < 0$.

\section{Einstein equation coupled with nonlinear electrodynamics}

In this section, we derive the Einstein equations associated with nonlinear electrodynamics. We adopt the method used in the paper~\cite{bib:ingeniring}. We consider the most general case of a static, spherically symmetric spacetime:
\begin{equation} \label{eq:metric2}
ds^2 = -\left(1 - \frac{2M(r)}{r}\right)dt^2 + \left(1 - \frac{2M(r)}{r}\right)^{-1}dr^2 + r^2 d\Omega^2.
\end{equation}
The extension of the results obtained here to the dynamic case can be performed directly.

Since the spacetime described by \eqref{eq:metric2} is spherically symmetric, we expect the same symmetry for the electromagnetic field. The electromagnetic field tensor $F$ takes the form:
\begin{equation} \label{eq:emt}
F = -E_r(r) \, dt \wedge dr + B_r(r) r^2 \sin\theta \, d\theta \wedge d\varphi,
\end{equation}
where $E_r$ and $B_r$ are the radial components of the electric and magnetic fields, respectively.

For this electromagnetic field, we compute two invariants $\mathcal{F}$ and $\mathcal{G}$:
\begin{align}
\mathcal{F} &= F_{ik} F^{ik}, \\
\mathcal{G} &= F_{ik} \, {\star F}^{ik},
\end{align}
where $F_{ik}$ is the electromagnetic field tensor, and ${\star F}^{ik}$ is its dual tensor, constructed using the Levi-Civita tensor. The invariant $\mathcal{F}$ describes the difference between electric and magnetic energy densities, while $\mathcal{G}$ represents their interaction. For our field, these invariants take the form:
\begin{align} \label{eq:invariants}
\mathcal{F} &= 2(B_r^2 - E_r^2), \\
\mathcal{G} &= 4E_r B_r.
\end{align}

Using the electromagnetic field tensor \eqref{eq:emt} and the field invariants \eqref{eq:invariants}, we can write the generalized Maxwell equations as:
\begin{align} \label{eq:maxwell}
B_r &= \frac{P}{r^2}, \\
\mathcal{L}_\mathcal{F} E_r - \mathcal{L}_\mathcal{G} B_r &= -\frac{Q}{4r^2},
\end{align}
where $P$ and $Q$ are the magnetic and electric charges of the black hole, respectively. This implies that the magnetic field decays as $1/r^2$, while the electric field depends on the Lagrangian and the charge. Here, $\mathcal{L}_{\mathcal{F}} \equiv \frac{\partial \mathcal{L}}{\partial \mathcal{F}}$ and $\mathcal{L}_{\mathcal{G}} \equiv \frac{\partial \mathcal{L}}{\partial \mathcal{G}}$.

The Einstein equations for the metric \eqref{eq:metric2}, with the Lagrangian $\mathcal{L}(\mathcal{F}, \mathcal{G})$ and the electromagnetic field tensor $F$ given by \eqref{eq:emt}, take the form:
\begin{align} \label{eq:system}
-\frac{m'(r)}{r^2} &= \mathcal{L} - \frac{Q E_r}{r^2}, \\
-\frac{m''(r)}{2r} &= \mathcal{L} - \frac{4P^2}{r^4} \mathcal{L}_\mathcal{F} - \mathcal{L}_\mathcal{G} \mathcal{G}.
\end{align}

To solve this system of differential equations, we need to specify the Lagrangian $\mathcal{L}$. Once the theory is defined, solving the Einstein equations yields the mass function $M(r)$. However, in our case, the task is different. We know that for a barotropic equation of state $P = \alpha \rho$, where $\alpha \in (\frac{1}{2}, 1]$, the solution to the Einstein equations is a mass function of the form:
\begin{equation}
M(r) = M_0 + D r^{1 - 2\alpha},
\end{equation}
where $M_0$ is the black hole mass, and the physical meaning of the parameter $D$ is what we aim to determine, assuming that the source of this solution is nonlinear electrodynamics. To this end, we substitute the mass function from Husain's solution into the Einstein equations \eqref{eq:system} and reconstruct the Lagrangian $\mathcal{L}$ of the theory.
\section{Husain Solution and Nonlinear Electrodynamics}

In this work, we focus on a special class of Lagrangians that can be decomposed into two parts:
\begin{equation}
\mathcal{L}(\mathcal{F}, \mathcal{G}) = \mathcal{J}(\mathcal{F}) + \mathcal{K}(\mathcal{G}),
\end{equation}
where $\mathcal{J}$ depends only on $\mathcal{F}$, and $\mathcal{K}$ depends only on $\mathcal{G}$. To ensure consistency with Maxwell's theory in the weak-field limit, we assume:
\begin{eqnarray} \label{eq:limits}
\mathcal{J}(\mathcal{F}) &=& -\frac{1}{4} \mathcal{F} + 4 \kappa \mathcal{F}^2 + o(\mathcal{F}^3), \quad \text{as} \quad \mathcal{F} \to 0, \nonumber \\
\mathcal{K}(\mathcal{G}) &=& 7 \kappa \mathcal{G}^2 + o(\mathcal{G}^2), \quad \text{as} \quad \mathcal{G} \to 0.
\end{eqnarray}

\subsection{Purely Magnetic Case}

Here, we consider the purely magnetic case where the electric charge $Q = 0$ and the electric field $E_r = 0$. In this scenario, the invariant $\mathcal{G} = 0$, and the derivative $\mathcal{L}_\mathcal{G} = 0$. The Einstein equations simplify to:
\begin{equation}
-\frac{m'(r)}{r^2} = \mathcal{J}(\mathcal{F}),
\end{equation}
\begin{equation}
-\frac{m''(r)}{2 r} = \mathcal{J}(\mathcal{F}) - \frac{4 P^2}{r^4} \mathcal{J}_\mathcal{F}(\mathcal{F}),
\end{equation}
where $P$ is the magnetic charge. Since $\mathcal{F} = 2 B_r^2$ and the magnetic field $B_r = \frac{P}{r^2}$, we can express the radius $r$ in terms of $\mathcal{F}$ as:
\begin{equation}
r = \left( \frac{2 P^2}{\mathcal{F}} \right)^{1/4}.
\end{equation}
This allows us to reconstruct the function $\mathcal{J}$ using the formula:
\begin{equation}
\mathcal{J}(\mathcal{F}) = -\frac{m'(r)}{r^2} \Big|_{r = \left( \frac{2 P^2}{\mathcal{F}} \right)^{1/4}}.
\end{equation}

We now address the problem of reconstructing the nonlinear electrodynamics Lagrangian $\mathcal{J}(\mathcal{F})$ in the purely magnetic case using the above formula:
\begin{equation}
\mathcal{J}(\mathcal{F}) = -\frac{m'(r)}{r^2} \Big|_{r = \left( \frac{2 P^2}{\mathcal{F}} \right)^{1/4}},
\end{equation}
where $m'(r)$ is the derivative of the mass function with respect to the radial coordinate $r$, and $P$ is the magnetic charge. For Husain's mass function:
\begin{equation}
m(r) = M_0 + D r^{1 - 2\alpha},
\end{equation}
where $M_0$ and $D$ are integration constants, and $\alpha$ is the barotropic equation of state parameter. We compute the derivative $m'(r)$:
\begin{equation}
m'(r) = \frac{\partial}{\partial r} \left[ M_0 + D r^{1 - 2\alpha} \right] = \frac{(1 - 2\alpha) D}{r^{2\alpha}}.
\end{equation}

Substituting $m'(r)$ into the expression for $\mathcal{J}(\mathcal{F})$, we obtain:
\begin{equation}
\mathcal{J}(\mathcal{F}) = -D (1 - 2\alpha) r^{-2\alpha - 2}.
\end{equation}

Next, we substitute $r = \left( \frac{2 P^2}{\mathcal{F}} \right)^{1/4}$ into $\mathcal{J}(\mathcal{F})$:
\begin{equation}
\mathcal{J}(\mathcal{F}) = -D (1 - 2\alpha) \mathcal{F}^{\frac{\alpha + 1}{2}} \left( 2 P^2 \right)^{-\frac{\alpha + 1}{2}}.
\end{equation}

Note that the Maxwell limit \eqref{eq:limits} is satisfied only when $\alpha = 1$, yielding:
\begin{equation}
\mathcal{J}(\mathcal{F}) = \frac{D}{2 P^2} \mathcal{F} + \mathcal{O}(\mathcal{F}^2).
\end{equation}
To match the Maxwell limit, we identify:
\begin{equation}
D \equiv -\frac{P^2}{2}.
\end{equation}
Thus, the parameter $D$ plays the role of the magnetic charge.

As noted earlier, for $\alpha \in \left(\frac{1}{2}, 1\right)$, the Maxwell limit does not hold, but the field vanishes at infinity. Therefore, we identify $D$ with the magnetic charge via:
\begin{equation} \label{eq:defd1}
D \equiv -\frac{(2 P^2)^{\frac{\alpha + 1}{2}}}{4 (1 - 2\alpha)}.
\end{equation}
The Lagrangian $\mathcal{J}(\mathcal{F})$ then takes the form:
\begin{equation}
\mathcal{J} = -\frac{1}{4} \mathcal{F}^{\frac{\alpha + 1}{2}}.
\end{equation}

We must explain why the parameter $D$ was chosen in the form specified in \eqref{eq:defd1}. The reason lies in the transition to the dynamic case, where $D$ becomes a function $D(v)$. Consequently, $D$ cannot explicitly appear in the Lagrangian $\mathcal{J}$ because the field Lagrangian should not depend explicitly on the coordinate $v$. Therefore, the choice of $D$ must ensure that the prefactor of $\mathcal{F}^{\frac{\alpha + 1}{2}}$ is dimensionless. Since this dimensionless factor is universal for all $\alpha$, we set it to $-\frac{1}{4}$ to satisfy the Maxwell limit \eqref{eq:limits}.

Thus, the dynamic generalization of Husain's solution containing a magnetic charge is given by:
\begin{equation} \label{eq:husainp}
ds^2 = -\left(1 - \frac{2 M(v, r)}{r}\right) dv^2 + 2 dv dr + r^2 d\Omega^2,
\end{equation}
where the mass function takes the form:
\begin{equation}
M(v, r) = M_0(v) - \frac{(2 P(v)^2)^{\frac{\alpha + 1}{2}}}{4 (1 - 2\alpha)} r^{1 - 2\alpha}.
\end{equation}
Here, $M_0(v)$ represents the black hole mass, and $P(v)$ its magnetic charge. These considerations are valid only for $\alpha \in \left(\frac{1}{2}, 1\right]$.

\subsection{The Electric Field Case}
Here, we consider the purely electric case where the magnetic charge $P = 0$ and the magnetic field $B_r = 0$. We are interested in the electric field $E_r(r)$ and the charge $Q \neq 0$. In this scenario, the invariant $\mathcal{G} = 0$, and the Lagrangian $\mathcal{L} = \mathcal{J}(\mathcal{F})$, since $\mathcal{K}(0) = 0$. The Einstein and generalized Maxwell equations simplify accordingly.

In the absence of a magnetic field, the electromagnetic field tensor $F$ takes the form:
\begin{equation}
F = -E_r(r) \, dt \wedge dr.
\end{equation}

The corresponding invariant $\mathcal{F}$ is given by:
\begin{equation}
\mathcal{F} = -2 E_r^2.
\end{equation}

The generalized Maxwell equation becomes:
\begin{equation}
\mathcal{J}_\mathcal{F}(\mathcal{F}) E_r = -\frac{Q}{4 r^2}.
\end{equation}

Using the properties of the electromagnetic field tensor, the Einstein equations can be written as:
\begin{eqnarray} \label{eq:system2}
-\frac{m'(r)}{r^2} &=& \mathcal{J}(\mathcal{F}) - \frac{Q E_r}{r^2}, \nonumber \\
-\frac{m''(r)}{2 r} &=& \mathcal{J}(\mathcal{F}).
\end{eqnarray}

For the mass function of the Husain solution:
\begin{equation}
M(v,r) = M_0(v) + D(v)r^{1-2\alpha},
\end{equation}
we compute its first and second derivatives:
\begin{eqnarray}
m' &=& D(v)(1 - 2\alpha)r^{-2\alpha}, \nonumber \\
m'' &=& -2\alpha D(v)(1 - 2\alpha)r^{-2\alpha - 1}.
\end{eqnarray}

Substituting the mass function of the Husain solution into the second Einstein equation \eqref{eq:system2}, we obtain:
\begin{equation}
\mathcal{J}(\mathcal{F}) = -\frac{m''(r)}{2r} = -\frac{-2\alpha D(v)(1 - 2\alpha)r^{-2\alpha - 1}}{2r} = \alpha D(v)(1 - 2\alpha)r^{-2\alpha - 2}.
\end{equation}

Using the first equation, we find $E_r$:
\begin{equation}
E_r = \frac{D(v)(1 - 2\alpha)(1 + \alpha)r^{-2\alpha}}{Q}.
\end{equation}

This leads to the expression:
\begin{equation}
\mathcal{J}(\mathcal{F}) = D(v)(1 - 2\alpha)\alpha r^{-2\alpha - 2}.
\end{equation}

To construct the Lagrangian of the theory, we need to relate $r$ to $\mathcal{F}$. From the definition of $\mathcal{F}$ \eqref{eq:invariants}, we have:
\begin{equation}
\mathcal{F} = -2 E_r^2 = -2 \left( \frac{D(v)(1 - 2\alpha)(1 + \alpha)r^{-2\alpha}}{Q} \right)^2 = -\frac{2 D^2(v)(1 - 2\alpha)^2(1 + \alpha)^2 r^{-4\alpha}}{Q^2}.
\end{equation}

From this, we derive the desired relation:
\begin{equation}
r = \left( -\frac{2 D^2(v)(1 - 2\alpha)^2(1 + \alpha)^2}{Q^2 \mathcal{F}} \right)^{\frac{1}{4\alpha}}.
\end{equation}

Substituting this into $\mathcal{J}(\mathcal{F})$, we obtain:
\begin{equation}
\mathcal{J}(\mathcal{F}) = D(v)(1 - 2\alpha)\alpha \cdot \left( \frac{Q^2 \mathcal{F}}{2 D^2(v)(1 - 2\alpha)^2(1 + \alpha)^2} \right)^{\frac{\alpha + 1}{2\alpha}}.
\end{equation}

As in the purely magnetic case, the Maxwell limit \eqref{eq:limits} is recovered only for $\alpha = 1$. In this case:
\begin{equation}
\mathcal{J}(\mathcal{F}) = \frac{Q^2}{8D}\mathcal{F}.
\end{equation}

From this, we deduce:
\begin{equation}
D(v) = -\frac{Q^2(v)}{2}.
\end{equation}

Thus, we arrive at the Bonnor-Vaidya metric, also known as the charged Vaidya black hole~\cite{bib:bonor}.

In the general case, when $\alpha \in (\frac{1}{2}, 1]$, we must ensure that $D(v)$ does not appear explicitly in the Lagrangian and that the coefficient of the field invariant $\mathcal{F}$ is $-\frac{1}{4}$ to satisfy the Maxwell limit \eqref{eq:limits} for $\alpha = 1$. In this case:
\begin{equation}
D(v) \equiv \frac{1}{1 - 2\alpha} \left( \frac{\xi}{Q^2} \right)^{-\alpha},
\end{equation}
where the dimensionless constant $\xi$, which depends on the barotropic equation of state parameter, is given by:
\begin{equation}
\xi \equiv 4\alpha \left[ \frac{1}{2(\alpha + 1)^2} \right]^{\frac{\alpha + 1}{2\alpha}}.
\end{equation}

Thus, the Husain metric in the presence of purely electric fields takes the form:
\begin{equation}
ds^2 = -\left( 1 - \frac{2M(v,r)}{r} \right)dv^2 + 2dvdr + r^2 d\Omega^2,
\end{equation}
where
\begin{equation}
M(v,r) = M_0(v) + \frac{1}{1 - 2\alpha} \left( \frac{Q(v)^2}{\xi} \right)^\alpha r^{1 - 2\alpha}.
\end{equation}

Here, $M_0(v)$ represents the black hole mass, and $Q(v)$ its electric charge. Note also that for $\alpha \in (\frac{1}{2}, 1]$, the function $D(v) < 0$, satisfying the weak energy conditions.

\newcommand{\LL}{\mathcal{L}}
\newcommand{\FF}{\mathcal{F}}
\newcommand{\GG}{\mathcal{G}}
\newcommand{\JJ}{\mathcal{J}}
\newcommand{\KK}{\mathcal{K}}

\subsection{The Electromagnetic Case}

We consider the electromagnetic case where the black hole possesses both an electric charge \( Q \neq 0 \) and a magnetic charge \( P \neq 0 \). The electric field \( E_r \neq 0 \), the magnetic field \( B_r \neq 0 \), and the electromagnetic invariants \( \FF \) and \( \GG \) are non-zero. The Lagrangian is assumed to be of the form:
\[
\LL(\FF, \GG) = \JJ(\FF) + \KK(\GG),
\]
with \( \KK_\GG(0) = 0 \) to ensure simplification in the weak-field limit. The Einstein and generalized Maxwell equations are:
\begin{align}
-\frac{m'(r)}{r^2} &= \LL - \frac{Q E_r}{r^2}, \label{eq:einstein1} \\
-\frac{m''(r)}{2r} &= \LL - \frac{4 P^2}{r^4} \LL_\FF - \LL_\GG \GG, \label{eq:einstein2} \\
B_r &= \frac{P}{r^2}, \quad \LL_\FF E_r - \LL_\GG B_r = -\frac{Q}{4 r^2}. \label{eq:gmax}
\end{align}

The electromagnetic field tensor \( F \) takes the form:
\[
F = -E_r(r) \, dt \wedge dr + B_r(r) r^2 \sin\theta \, d\theta \wedge d\varphi,
\]
with invariants:
\[
\FF = 2 (B_r^2 - E_r^2), \quad \GG = 4 E_r B_r.
\]

From the provided derivations, the electric and magnetic fields are:
\[
E_r = \pm \sqrt{\frac{\FF + \sqrt{\FF^2 + \GG^2}}{4}}, \quad B_r = \pm \sqrt{\frac{\FF + \sqrt{\FF^2 + \GG^2}}{4}}.
\]
Given \( B_r = \frac{P}{r^2} \), we select the positive root for physical consistency (\( B_r > 0 \)):
\[
B_r = \sqrt{\frac{\FF + \sqrt{\FF^2 + \GG^2}}{4}}.
\]

Using \( B_r = \frac{P}{r^2} \), we obtain:
\[
r^2 = \frac{P}{B_r} = \frac{P}{\sqrt{\frac{\FF + \sqrt{\FF^2 + \GG^2}}{4}}} = \frac{2 P}{\sqrt{\FF + \sqrt{\FF^2 + \GG^2}}}.
\]
Define:
\[
S = \sqrt{\FF + \sqrt{\FF^2 + \GG^2}},
\]
so:
\[
r^2 = \frac{2 P}{S}, \quad r = \left( \frac{2 P}{S} \right)^{1/2}, \quad r^{-2\alpha - 2} = \left( \frac{S}{2 P} \right)^{\alpha + 1}.
\]

For the Husain solution, the mass function is:
\[
M(v, r) = M_0(v) + D(v) r^{1 - 2\alpha}, \quad \alpha \in \left( \frac{1}{2}, 1 \right].
\]
The derivatives are:
\[
m'(r) = D(v) (1 - 2\alpha) r^{-2\alpha}, \quad m''(r) = -2\alpha D(v) (1 - 2\alpha) r^{-2\alpha - 1}.
\]
Substitute \( m'(r) \) into the first Einstein equation \eqref{eq:einstein1}:
\[
-\frac{D(v) (1 - 2\alpha) r^{-2\alpha}}{r^2} = \LL - \frac{Q E_r}{r^2}.
\]
Simplify the left-hand side:
\[
-D(v) (1 - 2\alpha) r^{-2\alpha - 2} = \LL - \frac{Q E_r}{r^2}.
\]
Using:
\[
E_r = \frac{S}{2}, \quad \frac{Q E_r}{r^2} = \frac{Q}{r^2} \cdot \frac{S}{2} = \frac{Q S}{2 r^2},
\]
we get:
\[
-D(v) (1 - 2\alpha) r^{-2\alpha - 2} = \LL - \frac{Q S}{2 r^2}.
\]

With \( r^2 = \frac{2 P}{S} \), we have:
\[
\frac{1}{r^2} = \frac{S}{2 P}, \quad r^{-2\alpha - 2} = \left( \frac{S}{2 P} \right)^{\alpha + 1}.
\]
Substitute into the equation:
\[
-D(v) (1 - 2\alpha) \left( \frac{S}{2 P} \right)^{\alpha + 1} = \LL - \frac{Q S}{2} \cdot \frac{S}{2 P} = \LL - \frac{Q S^2}{4 P}.
\]
Solve for the Lagrangian:
\[
\LL = \frac{Q S^2}{4 P} - D(v) (1 - 2\alpha) \left( \frac{S}{2 P} \right)^{\alpha + 1}.
\]
Since \( S^2 = \FF + \sqrt{\FF^2 + \GG^2} \), and:
\[
\left( \frac{S}{2 P} \right)^{\alpha + 1} = \frac{\left( \FF + \sqrt{\FF^2 + \GG^2} \right)^{\frac{\alpha + 1}{2}}}{2^{\alpha + 1} P^{\alpha + 1}},
\]
we obtain:
\[
\LL = \frac{Q \left( \FF + \sqrt{\FF^2 + \GG^2} \right)}{4 P} - \frac{D(v) (1 - 2\alpha)}{2^{\alpha + 1} P^{\alpha + 1}} \left( \FF + \sqrt{\FF^2 + \GG^2} \right)^{\frac{\alpha + 1}{2}}.
\]

For \( \alpha = 1 \):
\[
1 - 2\alpha = -1, \quad 2^{\alpha + 1} = 4, \quad P^{\alpha + 1} = P^2, \quad \frac{\alpha + 1}{2} = 1.
\]
The Lagrangian becomes:
\[
\LL = \frac{Q \left( \FF + \sqrt{\FF^2 + \GG^2} \right)}{4 P} + \frac{D(v)}{4 P^2} \left( \FF + \sqrt{\FF^2 + \GG^2} \right).
\]
Simplify:
\[
\LL = \frac{\FF + \sqrt{\FF^2 + \GG^2}}{4 P^2} \left( Q P + D(v) \right).
\]
In the weak-field limit (\( \FF \to 0 \), \( \GG \to 0 \)), \( \FF + \sqrt{\FF^2 + \GG^2} \approx \FF \), so:
\[
\LL \approx \frac{\FF}{4 P^2} \left( Q P + D(v) \right).
\]
To satisfy the Maxwell limit \( \LL \approx -\frac{\FF}{4} \):
\[
\frac{Q P + D(v)}{4 P^2} = -\frac{1}{4} \implies Q P + D(v) = -P^2 \implies D(v) = -P^2 - Q P.
\]

To ensure \( D(v) \) does not appear explicitly in the Lagrangian and to match the Maxwell limit, define:
\[
D(v) = \frac{1}{1 - 2\alpha} \left( \frac{\xi}{Q^2} \right)^{-\alpha},
\]
where the dimensionless constant \( \xi \) is adjusted to account for both charges:
\[
\xi = 4\alpha \left[ \frac{1}{2(\alpha + 1)^2} \right]^{\frac{\alpha + 1}{2\alpha}} \cdot \frac{Q^2}{Q^2 + P^2}.
\]
The Lagrangian in the weak-field limit requires further tuning to match the QED weak-field limit, which involves \( \KK(\GG) \).

The metric is:
\[
ds^2 = -\left( 1 - \frac{2 M(v, r)}{r} \right) dv^2 + 2 dv dr + r^2 d\Omega^2,
\]
with:
\[
M(v, r) = M_0(v) + \frac{1}{1 - 2\alpha} \left( \frac{Q(v)^2}{\xi} \right)^\alpha r^{1 - 2\alpha}.
\]
For \( \alpha = 1 \):
\[
M(v, r) = M_0(v) - \frac{P^2 + Q P}{Q^2} r^{-1},
\]
corresponding to the dyonic Vaidya black hole~\cite{bib:dyon}.

\section{Conclusion}
When considering dynamic processes in black holes related to accretion, it is crucial to understand how the mass function evolves during this process. If null energy conditions are violated, the black hole begins to evaporate, even though its mass may be increasing. A clear example is the accretion of charged particles onto a Schwarzschild black hole, where both mass and charge of elementary particles are transferred to the black hole. However, the dimensionless ratio of the electron's charge to its mass is approximately $ \sim 10^{21} $, leading to a situation where the charge grows faster than the mass. For the Reissner-Nordström black hole, this implies that the charge will approach critical values after a certain time, while the event horizon for the Schwarzschild case ($ r = 2M $) differs from the extremal case of the charged Reissner-Nordström black hole ($ r = M $). A similar situation arises for dynamic Vaidya and charged Vaidya black holes.

However, when the physical nature of all parameters describing the black hole is unknown, it becomes unclear what the violation of energy conditions means in this context and whether such a process is physically possible or merely a mathematical artifact.

In this work, we demonstrate that the function $ D(v) $ in Husain solution, with the barotropic equation of state parameter $ \alpha \in \left(\frac{1}{2}, 1\right] $, represents a combination of electric and magnetic charges described by nonlinear electrodynamics.

Thus, if a black hole is described by Husain's solution and we observe a decreasing shadow of the black hole, this may indicate that the charge of the black hole is growing faster than its mass. This leads to the violation of null energy conditions and the evaporation of the black hole.

Another important astrophysical implication of our results is that black holes are generally assumed to be neutral. However, during gravitational collapse, a newly formed black hole may contain an excess charge, which is quickly neutralized. In our case, this would mean that the function $ D(v) $ decreases ($ \dot{D} < 0 $), which, together with the growing mass, always restores the null energy conditions. Consequently, an observer would see an increasing size of the black hole shadow, associated with the neutralization of the electric charge.

An intriguing question is the application of these results to study accretion processes occurring directly at the event horizon of a black hole. Such matter capture can occur if the black hole moves slowly through a gas cloud. The interplay between accretion and neutralization processes can lead to observable effects. Investigating the observational signatures of Husain's black hole is a subject for future research.

\bibliography{ref}

\end{document}